\documentclass[conference]{IEEEtran}

\usepackage{color}

\def\ignore#1{}

\usepackage[normalem]{ulem}
\usepackage{bm}

\IEEEoverridecommandlockouts
\usepackage{confs}
\usepackage{xspace}
\usepackage{booktabs}
\usepackage{cite}
\usepackage{subcaption}

\usepackage{stfloats}

\def\Agent{{\it Agent}}
\def\Agents{{\it Agents}}
\def\AgentSet{\mathcal{I}} \def\ActionSet{\mathcal{A}}
\def\States{\mathcal{S}} \def\Obs{\Omega}
\def\PosInteger{\mathbb{Z}^+}
\def\DCset{\mathcal{D}}
\def\ADC{M^{\it ADC}}
\def\Env{\mathcal{E}}

\begin{document}

\title{Strategy-Following Multi-Agent Deep Reinforcement Learning Considering Control Strategies toward Other Agents
}

\title{Strategy-Following Multi-Agent Deep Reinforcement Learning Considering Control Strategies Provided to Other Agents
}

\author{\IEEEauthorblockN{Yamato Takahagi\IEEEauthorrefmark{1},
    Gentoku Nakasone\IEEEauthorrefmark{1}, Yoshinari Motokawa\IEEEauthorrefmark{1}
    and Toshiharu Sugawara\IEEEauthorrefmark{2}\thanks{This work is partly supported by KAKENHI 25K03188.}}
	\IEEEauthorblockA{\textit{Department of Computer Science and Communications Engineering}\\
	  \textit{Waseda University}, Tokyo 1698555, Japan\\
	  Email: \IEEEauthorrefmark{1}\{y.takahagi,g.nakasone,y.motokawa\}@isl.cs.waseda.ac.jp,
	  \IEEEauthorrefmark{2}sugawara@waseda.jp}}

\maketitle

\begin{abstract}
This study proposes a learning method for multi-agent systems that
allows agents to be controlled through human manager instructions
after learning and enables uninstructed agents to implicitly
complement the overall work based on the actions of other
agents. Multi-agent applications using deep learning have shown
potential; thus, to achieve extensive social applications, humans
should be able to control learned agents using simple methods to
respond to environmental and social changes. Even without such
changes, learned coordination often does not match the expectations of
human managers, making it preferable to control coordination
structures to match human intentions. Some studies have aimed to
control agent behavior using simple instructions. However, they
assumed that instructions are provided to all agents, which is
time-consuming and not evident when designing a better cooperation
regime. Ideally, specific agents should receive key action
instructions, while others should automatically complete the remaining
tasks. The proposed method, which extends previous work on
controllability in multi-agent deep reinforcement learning, enables
uninstructed agents to adaptively complement overlooked tasks and
areas. The experimental results show that agents using the proposed
method can shift to another cooperative structure and achieve better
performance than those using conventional methods.
\end{abstract}


\section{Introduction}
Recent advances in artificial intelligence (AI) have led to the widespread use of multiple AI agents. As these agents become more prevalent, their interactions also increase. Therefore, research on multi-agent systems that enable appropriate coordination and conflict avoidance between agents has received attention. However, the coordination and collaboration structures between agents remain complex and challenging. Recent advancements in {\em deep reinforcement learning} (DRL) have resulted in various studies on {\em multi-agent DRL} (MADRL) to establish coordinated behavior using environmental data. For example, it has been utilized in autonomous driving~\cite{shalev2016safe} and real-time strategic games~\cite{samvelyan2019starcraft}. They are also utilized to optimize communication among UAVs, ground vehicles, and portable devices~\cite{shi2024task,zhang2024joint}.
\par

Although MADRL can facilitate the formation of cooperative structures, several challenges remain. These include the {\em lazy-agent problem}, where agents learn that inaction is advantageous because of the effective performance of other agents, and the {\em unaligned coordination structure}, where cooperation converges to a local optimum that may not align with the objectives of human managers. Even if the initial cooperative structure is suitable, changes in requirements or environmental conditions may necessitate non-local adjustments to the coordinated behavior. However, MADRL requires extensive training periods even for minor modifications, which makes retraining for each change inefficient. One potential solution is to integrate the controllability of human managers into the learning process, thereby allowing agents to follow their instructions/directions.
\par

Several studies have explored the establishment of controllability within a DRL framework for harmonious machine-human coexistence~\cite{winograd1972understanding}. For example, research on {\em reinforcement learning} (RL) has focused on controllability enhancement through linguistic instructions to direct agent behavior~\cite{fried2018speaker,misra2018mapping}. DRL studies have proposed language-based command methods~\cite{bahdanau2018learning} and approaches such as {\em controllable imitative RL} (CIRL)~\cite{liang2018cirl}, which incorporates controllability into the policy design of DRL agents in autonomous driving. Studies have also explored the control of learned agents using programmed instructions~\cite{sun2020program,chen2020ask,yang2021program,gangopadhyay2021hierarchical}. However, most studies have assumed control of a single agent. Because coordinated behavior between agents requires sophisticated analysis, these methods cannot simply be extended to MADRL.
\par

Motokawa and Sugawara~\cite{motokawa2023strategy} proposed a {\em strategy-following distributed attentional actor architecture after conditional attention} (sfDA6-X) to establish controllability in MADRL. As a key feature, sfDA6-X distributes abstract directions from an external manager to agents rather than using detailed programmatic or linguistic commands, and the agents autonomously determine behaviors based on policies learned in several directions. Because actions are generated using learned policies, fine-grained instructions are not required, allowing the system to naturally cover the environment cooperatively. The use of explicit directions during learning helps prevent lazy agents.
\par

However, as the number of agents increases, providing directions for all agents becomes difficult, potentially causing behaviors that are misaligned with environmental structures. When instructions are given only to certain agents, imbalances may occur, leading to inadequate environmental coverage. Uninstructed agents must act flexibly and autonomously in ways that complement cooperative work by considering directions given to others.
\par

To address this issue, we propose a method that enables agents to consider the directions given to others and autonomously determine their behaviors, allowing them to complement each other without compromising efficiency. Specifically, we propose a learning method by which agents receive aggregated directional information from others, ensuring their independence from the agent numbers. We experimentally demonstrated that our method encourages cooperative behavior by considering others' actions based on directions using the {\em object collection game} employed in~\cite{motokawa2023strategy}. We analyzed the resulting agent behaviors to understand the performance of the proposed method.
\par

\section{Related Work}
Several studies have aimed to improve agent learning efficiency by allowing human instruction during training. Ross et al.~\cite{ross2011reduction} proposed a method that involves humans controlling agents in real time, enabling better imitation of rational behavior. Wu et al.~\cite{wu2023toward} improved learning in autonomous driving by allowing humans to correct inappropriate agent actions. Rajeswaran et al.~\cite{rajeswaran2016epopt} proposed treating human-selected actions as agent actions and updating the policy functions based on rewards. Although these approaches enhance learning efficiency, they do not address the control of agent behavior after training. Our study focused on establishing agent controllability after training.
\par

Some studies on DRL have focused on the continual controllability of agents even after the training phase. Fried et al.~\cite{fried2018speaker} realized vision-and-language navigation using embedded speaker models and {\em language-conditioned image generation networks}. Other studies have demonstrated that agents can learn to follow instructions written in formal languages, enabling them to exhibit behaviors that align with human intentions~\cite{sun2020program,chen2020ask,yang2021program,gangopadhyay2021hierarchical}. However, these studies mainly focused on single-agent settings and did not address multi-agent systems.
\par

In contrast, Motokawa et al.~\cite{motokawa2023interpretability} proposed sfDA6-X, a method for controlling post-training agents in multi-agent environments using rough strategies from human managers. This approach enables agents to acquire flexible behaviors and form coordinated actions based on human instructions. However, sfDA6-X has a limitation: when strategic directions from humans are biased, agents may act independently without considering the directions provided to other agents, thereby reducing efficiency. Therefore, we aimed to improve agent’s performance by developing a learning method by which each agent considers the strategic directions given to the other agents. This approach seeks to ensure controllability while enabling agents to complement each other efficiently.
\par

\section{Preliminaries}
\subsection{Decentralized POMDP}
We consider discrete time $t$ ($\in\PosInteger$) in units of {\em timestep} or simply {\em step}, where $\PosInteger$ is the set of positive integers. Let $\AgentSet = \{1, \ldots, N\}$ denote the set of agents. Our framework is based on a {\em decentralized partially observable Markov decision process} (Dec-POMDP)~\cite{puterman2014markov} for $N$ agents. A Dec-POMDP is represented by a tuple $\langle\AgentSet, \States, \{\ActionSet_i\}, p_T, \{r_i\}, \{\Obs_i\}, \mathcal{O}, H\rangle$, where $\States$ is the set of environment states, and $\ActionSet_i$ is the set of actions available to agent $i \in \AgentSet$, with the set of joint actions $\ActionSet = \ActionSet_1\times\dots\times\ActionSet_N$. Given the joint action $a \in \ActionSet$ and states $s, s' \in \States$, $p_T(s'|s,a)\in[0, 1]$ denotes the {\em state transition function}, and $r_i(s,a) \in \mathbb{R}$ denotes the {\em reward function} for agent $i$. Each $\Obs_i$ is the finite set of possible observations for agent $i$. Given a joint observation $o \in \Obs = \Obs_1 \times \dots \times \Obs_N$, $\mathcal{O}(o|s,a)$ represents the conditional observation probability function. $H\in\PosInteger$ denotes the maximum number of steps in the stochastic process. Each agent $i \in \AgentSet$ learns a policy $\pi_i$ to maximize its own {\em discounted cumulative reward} $R_i = \sum_{t=0}^H \gamma^t r_i(s,a)$. Here, $0 \leq \gamma < 1$ is the {\em discount factor}, which determines the present value of future rewards. A higher value of $\gamma$ indicates that future rewards have a greater influence.
\par

\begin{figure}[t]
  \centering
  \includegraphics[keepaspectratio, width=0.8\linewidth]{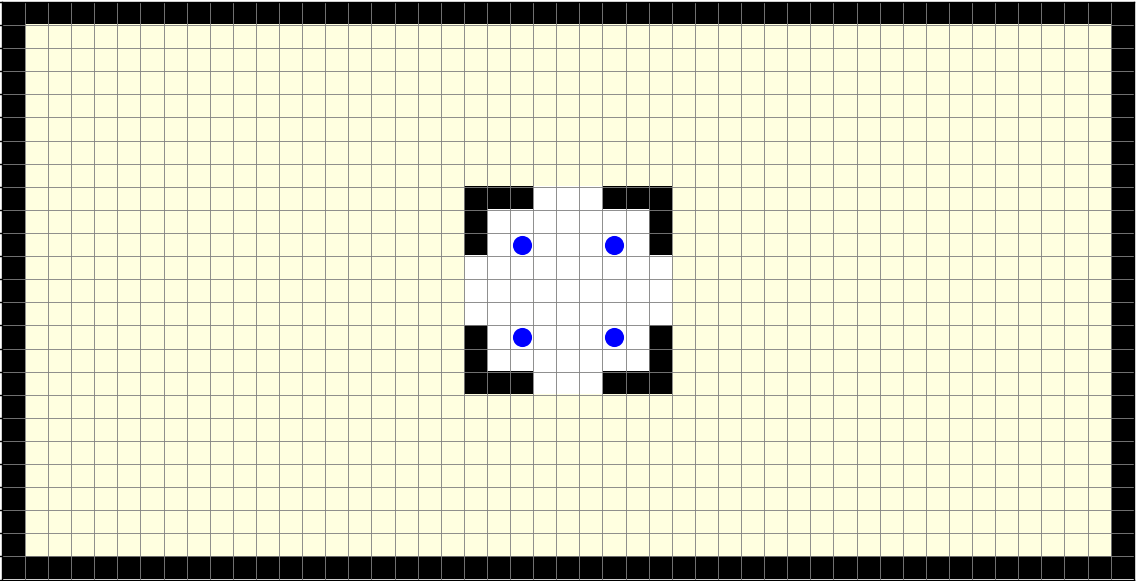}
  \caption{Experimental environment}
  \label{fig:map}
\end{figure}

\subsection{Object Collection Game}
An object collection game in a simple $G_X \times G_Y$ grid environment is used for the evaluation. In this game, the agents start moving at time $t=0$ from predetermined initial positions (blue nodes). The objects initially spawn in the environment. Agent $i$ selects action $a_i \in \ActionSet_i = \{ {\it up, down, left, right}\} $ at each step. If agent $i$ moves to a node with an object, it collects the object and receives a positive reward, $r_o > 0$. The collected object then disappears, and another spawns elsewhere. If agent $i$ collides with a wall or another agent, it receives a negative reward $r_c < 0$. Otherwise, it receives a small negative reward $r_m$ ($\leq 0$) as a movement cost. The agents aim to collect objects in designated areas (yellow areas) while avoiding collisions with walls (black nodes) and other agents.
\par

\subsection{Observation Matrices Based on Local View and Global Position}
The observation of an agent can be expressed using two types of matrices~\cite{miyashita2021analysis}. The first type, called {\em local observation matrices} (or {\em local matrices}), is generated from a local visible area of size $R_X\times R_Y$ centered on $i$, where $R_X$ and $R_Y$ are odd integers such that $3\leq R_X\leq G_X$ and $3\leq R_Y\leq G_Y$. The local information is encoded into $N_c$ channels of $R_X \times R_Y$ matrices (a tensor of shape $N_c \times R_X \times R_Y$), representing observable items such as agents (with the agent centered), obstacles, and objects being collected and input into each agent's network; thus, $N_c=N+2$. The second matrix, called the {\em positional matrix}, represents $i$'s global position in the environment. A $G_X \times G_Y$ matrix represents the entire environment, where the element at $i$'s position is $1$ and the others are $0$. Agents can obtain their global positions using GPS.
\par

\subsection{Multi-Head Attention}
The self-attention mechanism~\cite{vaswani2017attention} calculates the similarity between the vectors in a matrix using the following equation:
\begin{equation}\notag
  \mathrm{Attention}(Q,K,V) = \mathrm{Softmax}(\frac{Q\cdot K^T}{\sqrt{d}})V,
\end{equation}
where $Q$, $K$, and $V$ represent the query, key, and value matrices, respectively, and $d$ denotes the dimensionality of the query/key vectors. Furthermore, multi-head attention applies the self-attention mechanism in parallel across $h$ ($h\in\PosInteger$) attention heads and is computed as follows:
\begin{align}
  \mathrm{MHA}(Q, K, V) &= \mathrm{Concat}(\mathit{head}_1,
	\ldots, \mathit{head}_h)W^O \notag \\
  \mathit{head}_l &= \mathrm{Attention}(Q \cdot W_l^Q, K \cdot W_l^K, V \cdot W_l^V),
  \notag
\end{align}
where $W_l^Q$, $W_l^K$, $W_l^V$, and $W^O$ are learnable parameter matrices for $\mathit{head}_l$ ($1 \leq \forall l \leq h$).
\par

\begin{figure}[t]
  \begin{minipage}{0.22\linewidth}
      \centering
      \includegraphics[width=\linewidth]{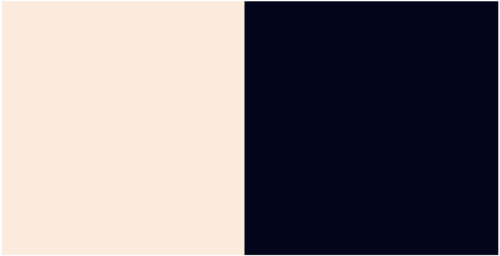}
  \end{minipage}
  \hfill
  \begin{minipage}{0.22\linewidth}
      \centering
      \includegraphics[width=\linewidth]{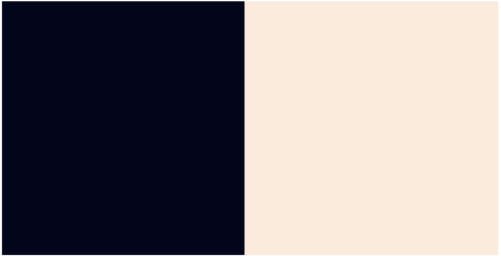}
  \end{minipage}
  \hfill
  \begin{minipage}{0.22\linewidth}
      \centering
      \includegraphics[width=\linewidth]{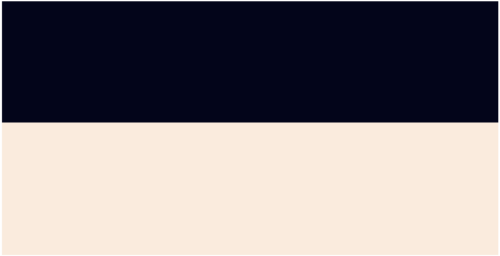}
  \end{minipage}
  \hfill
  \begin{minipage}{0.22\linewidth}
      \centering
      \includegraphics[width=\linewidth]{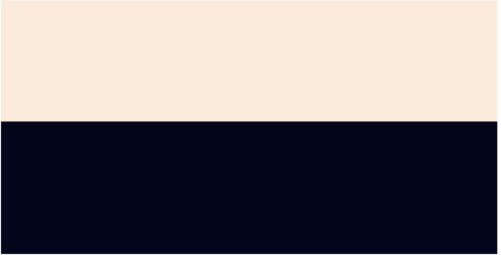}
  \end{minipage}
  \caption{Example of destination channels}
  \label{fig:DCex}
\end{figure}

\subsection{Destination Channels}
Here, we describe {\em destination channels} (DCs)~\cite{motokawa2023strategy} that direct agents to perform actions. DCs are the strategy directions provided to agents as matrices. They do not contain detailed instructions for the agents to follow. A DC is a matrix of size $G_X \times G_Y$ with binary values (0 and 1) as elements. Figure~\ref{fig:DCex} shows four examples of DCs in the environment shown in Fig.~\ref{fig:map}, where the beige regions correspond to the elements filled with $1$. During training, the network in each agent receives observation matrices and one DC and learns to obtain a positive reward $r_i(s,a) = r_o$ only when collecting an object in the beige region. However, a reward cannot be obtained if an object is collected from another region. By preparing the necessary DC shapes and providing DCs during the training and execution phases, the agent's behaviors can be controlled with a certain granularity of control to work in desirable areas.
\par

\begin{figure}[t]
  \centering
  \begin{minipage}[t]{0.49\linewidth}
    \centering
    \includegraphics[keepaspectratio, height=5.4cm]{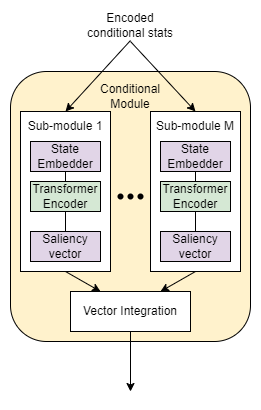}
  \end{minipage}
  \begin{minipage}[t]{0.49\linewidth}
    \centering
    \includegraphics[keepaspectratio, height=5.4cm]{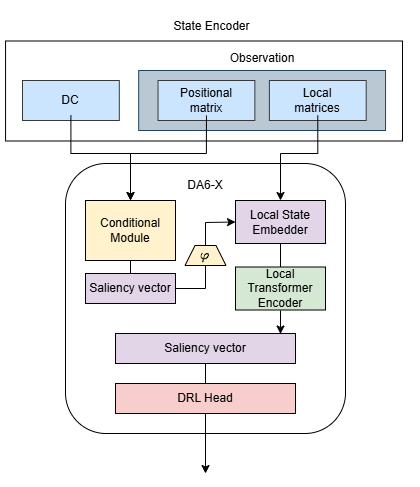}
  \end{minipage}
  \caption{Architecture of sfDA6-X}
  \label{fig:DA6-X}
\end{figure}

\subsection{sfDA6-X}
We describe sfDA6-X~\cite{motokawa2023strategy}, which forms the basis of our study. sfDA6-X is an extended model based on its predecessor, DA6-X~\cite{motokawa2023interpretability}. As shown in Fig.~\ref{fig:DA6-X}, DA6-X consists of three stages: a conditional module (CM), local transformer encoder, and DRL head. The CM recognizes environmental conditions (specified by the positional matrix in the object collection game; agents can learn rules depending on the position), and the local transformer encoder assigns weights to the observed information through an attention mechanism. In sfDA6-X, the CM is extended to receive environmental conditions and strategic directions as DCs, thereby enabling model recognition. Using the transformer encoder's self-attention mechanism in the CM, the model generates a saliency vector representing the elements most attended to during decision-making. This saliency vector is reused in the local transformer encoder and DRL head. The DRL head outputs agent actions based on the input features, corresponding to "X" in the method name.
\par

Assuming the existence of $M$ submodules, the $m$-th module $(1 \leq m \leq M)$ processes the patched conditional state matrix $x_m \in \mathbb{R}^{I_m \times (P^2_m N_m)}$ and embeds it using the embedding parameter matrix $E^\mathrm{cond}_m \in \mathbb{R}^{(P^2_m N_m) \times C_m}$, where $I_m$ is the length of the $m$-th conditional state after patching, $P_m$ is the patch size, $N_m$ is the number of channels before patching, and $C_m$ is the dimensionality of the $m$-th saliency vector. The embeddings are concatenated with the saliency vector $\bm{u}^{\mathrm{sal}}_m \in \mathbb{R}^{C_m}$ and reused from the previous stage. This operation is represented as follows: 
\begin{equation}\notag
  g_{m,0} = [\bm{u}^{\mathrm{sal}}_m; x^1_m E^\mathrm{cond}_m; x^2_m E^\mathrm{cond}_m; \ldots ; x^{I_m}_m E^\mathrm{cond}_m] + P^{\mathrm{cond}}_m,
\end{equation}
where $P^{\mathrm{cond}}_m \in \mathbb{R}^{(I_m + 1) \times C_m}$ denotes the positional embedding corresponding to the state-embedding operation in the CM submodules, as shown in Fig. ~\ref{fig:DA6-X}.
\par

After this operation, the resulting data are fed into the transformer encoder within the submodule. Letting the transformer encoder be denoted by $\mathrm{TEL}^\mathrm{cond}_m$, this operation is represented by the following equation: 
\begin{equation}\notag
  g_{m,l} = \mathrm{TEL}^\mathrm{cond}_m (g_{m, l-1}), l = 1, \ldots , L_m.
\end{equation}
After repeating this process $L_m$ times, the saliency vectors obtained from all submodules ($g^0_{1,L_1}, \ldots, g^0_{m,L_m}, \ldots, g^0_{M,L_M}$) are aggregated by vector integration to generate the final saliency vector $\bm{v}^{\mathrm{sal}} \in \mathbb{R}^{C_\mathrm{L}}$ of length $C_\mathrm{L}$. This operation is expressed as follows:
\begin{equation}\notag
  \bm{v}^{\mathrm{sal}} = \mathrm{VectorIntegration}(g^0_{1,L_1}, \ldots, g^0_{m,L_m}, \ldots, g^0_{M,L_M}).
\end{equation}
Through this integration process, the information from all submodules is consolidated into a unified feature representation. The generated $\bm{v}^{\mathrm{sal}}$ is then updated and passed to the DRL head as follows:
\begin{equation}\notag
  \begin{split}
    h_0 &= [\varphi(\bm{v}^{\mathrm{sal}}); y^1 E^\mathrm{local}; y^2 E^\mathrm{local}; \ldots ; y^{I_\mathrm{L}} E^\mathrm{local}] + P^{\mathrm{local}}\\
    h_l &= \mathrm{TEL}^{\mathrm{local}}(h_{l-1}), \qquad l = 1, \ldots , L_\mathrm{L}\\
    Q &= \mathrm{DRLHead}(h^0_{L_\mathrm{L}}),
  \end{split}
\end{equation}
where $\varphi(\cdot)$ is the projection operation, and $\bm{v}$ is passed to the local state embedder (Fig.~\ref{fig:DA6-X}). Subsequently, the local observation $y \in \mathbb{R}^{I_\mathrm{L} \times (P^2_\mathrm{L} N_\mathrm{L})}$ is embedded using the parameter $E_\mathrm{L} \in \mathbb{R}^{(P^2_\mathrm{L} N_\mathrm{L}) \times C_\mathrm{L}}$. The embedding result is then concatenated with $\varphi(\bm{v}^{\mathrm{sal}})$, where $I_L$ is the local observation length after patching, $P_L$ is the local patch size, and $N_L$ is the number of channels in the patched local observation.
\par

Subsequently, the input $h_0$ is passed through the local transformer encoder layers, denoted by $\mathrm{TEL}^{\mathrm{local}}$, for $L_\mathrm{L}$ iterations ($L_\mathrm{L} \in \PosInteger$), and the positional embedding $P^{\mathrm{local}} \in \mathbb{R}^{(I_\mathrm{L}+1)\times C_\mathrm{L}}$ is added. Finally, the generated saliency vector $h^0_{L_\mathrm{L}}$ is fed into the DRL head.
\par

Therefore, the sfDA6-X model integrates self-attention, enhancing action selection and enabling efficient cooperative exploration. By reusing the saliency vector from the CM, the agent learns policies by considering the environmental conditions and strategy directions from the DC, thereby ensuring controllability.
\par

\begin{figure}[t]
  \centering
  \includegraphics[keepaspectratio, width=0.8\linewidth]{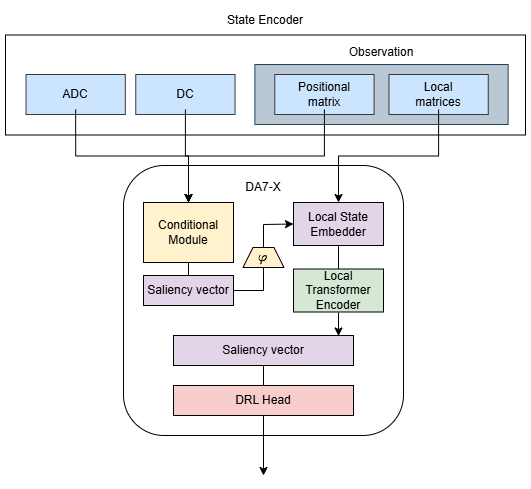}
  \caption{Architecture of sfDA7-X}
  \label{fig:DA7-X}
\end{figure}

\section{Proposed Method}
We introduce a method called {\em strategy-following distributed attentional actor architecture after conditional attention augmented by aggregated DC} (sfDA7-X), which is an extension of sfDA6-X. In this method, an agent not only receives its own direction (i.e., DC) but also incorporates an abstracted representation of the DCs assigned to other agents, enabling it to infer their behavior while making its own decisions. For example, even when the agent does not receive explicit instructions (or is instructed with no constraints), it aims to learn a cooperative policy that covers the overall task while minimizing redundancy based on the directions given to other agents. Because the number of other agents is not fixed, we generate an abstracted representation called an {\em aggregated DC} (ADC) by encoding the DCs assigned to all other agents into a single DC. The generated ADC is fed into the network as an additional input.
\par

The ADC is generated as follows. Let the set of all DCs be denoted as $\DCset$. During training, for each episode, a DC randomly selected from $\DCset$ is assigned to each agent. Therefore, the agents usually have different DCs. In episode $e$, let the set of all DCs assigned to agents other than $i \in \AgentSet$ be denoted by $\mathcal{D}_{e,i}$. The ADC assigned to agent $i$ ($\ADC_i$) is defined as
\begin{equation}
  M^{\it ADC}_i = \sum_{d \in \DCset_{e,i}} d.
\end{equation}
Notably, the ADC is represented as a single matrix, regardless of the number of agents $N$. Through this operation, the exploration strategy of the agent with respect to its destination is updated in harmony with the exploration areas of other agents and is expected to efficiently collect objects while avoiding redundancy. Fig.~\ref{fig:DA7-X} shows the architecture of the proposed sfDA7-X model. Note that we assume that agents can obtain the DCs of others during the execution phase; we believe that this is acceptable because DCs are determined and provided by human managers. Unlike sfDA6-X, which only considers the DC assigned to the agent itself, sfDA7-X incorporates the ADC, enabling uninstructed agents to complement overlooked areas.
\par

\begin{figure}[t]
  \centering
  \begin{subfigure}{0.19\linewidth}
    \centering
    \includegraphics[width=\linewidth]{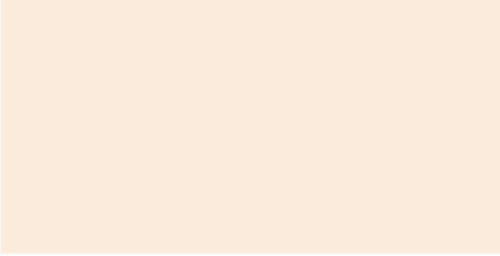}
  \end{subfigure}
  \begin{subfigure}{0.19\linewidth}
      \centering
      \includegraphics[width=\linewidth]{figures/half0.PNG}
  \end{subfigure}
  \begin{subfigure}{0.19\linewidth}
      \centering
      \includegraphics[width=\linewidth]{figures/half1.PNG}
  \end{subfigure}
  \begin{subfigure}{0.19\linewidth}
      \centering
      \includegraphics[width=\linewidth]{figures/half2.PNG}
  \end{subfigure}
  \begin{subfigure}{0.19\linewidth}
      \centering
      \includegraphics[width=\linewidth]{figures/half3.PNG}
  \end{subfigure}

  \caption{DCs assigned during training}
  \label{fig:DCexp}
\end{figure}

\begin{table}[t]
  \caption{Training hyperparameters}\label{table:1}
  \centering
  \begin{tabular}[t]{ll|ll}
    \toprule
    Hyperparameter & Value & Hyperparameter & Value \\
    \cmidrule(rr){1-2}\cmidrule(ll){3-4}
      Learning rate & $1\mathrm{e}^{-3}$ & Discount factor $\gamma$ & $0.9$ \\
      Adam $\varepsilon$ & $1\mathrm{e}^{-8}$ & Batch size & $32$ \\
       \bottomrule
  \end{tabular}
\end{table}

\section{Experimental Evaluation}
\subsection{Experimental Settings}
We experimentally evaluated the proposed method using an object collection game in a grid environment $\Env$ of size $G_X \times G_Y = 49 \times 25$, as shown in Fig.~\ref{fig:map}, where black areas represent walls, blue nodes indicate the initial positions of agents, white areas denote regions without objects, and yellow areas indicate regions where objects can appear. The number of agents was $N=4$. At the start of the episode, each agent was placed at a predetermined position (blue node) and assigned one direction (DC) that was randomly selected from the set $\DCset$ during training. The experiment was limited to two types of DCs: {\em entire} (i.e., agents that can collect objects anywhere) and {\em half} DCs, as illustrated in Fig.~\ref{fig:DCexp}. In the proposed method, each agent receives an ADC and DC. The environment maintained $80$ objects, and the collected objects were regenerated at random, as shown by empty yellow nodes. The episode length was $H = 200$ steps, and the training period was $20,000$ episodes ($4$M steps). The baseline was sfDA6-X because the other methods cannot accept DCs. A comparison with other methods that do not employ DCs can be found in \cite{motokawa2023strategy}. We analyzed objects collected during training and working areas, where agents collected objects during execution when DCs were provided for control, and compared the results with those of the baseline.
\par

In the execution phases, we examined the performance when biased DCs that directed agents to collect objects, mainly in a specific area, were provided. Our concern is how agents without specific directions, i.e., agents with entire DCs, can cover other areas in accordance with the game's meaning. We set $r_o=1$, $r_c=-1$, and $r_m=0$. The agent's field of view was set to $R_X=R_Y= 7$. We employed the {\em implicit quantile network} (IQN)~\cite{dabney2018implicit} for the DRL heads (denoted by sfDA6-IQN and sfDA7-IQN). The hyperparameters used in training are listed in Table~\ref{table:1}. The greedy parameter $\varepsilon$ was set to $1.0$ and decreased to $0.05$ with a decay rate of $0.998$.
\par

\begin{figure}[t]
  \centering
  \includegraphics[width=0.8\linewidth]{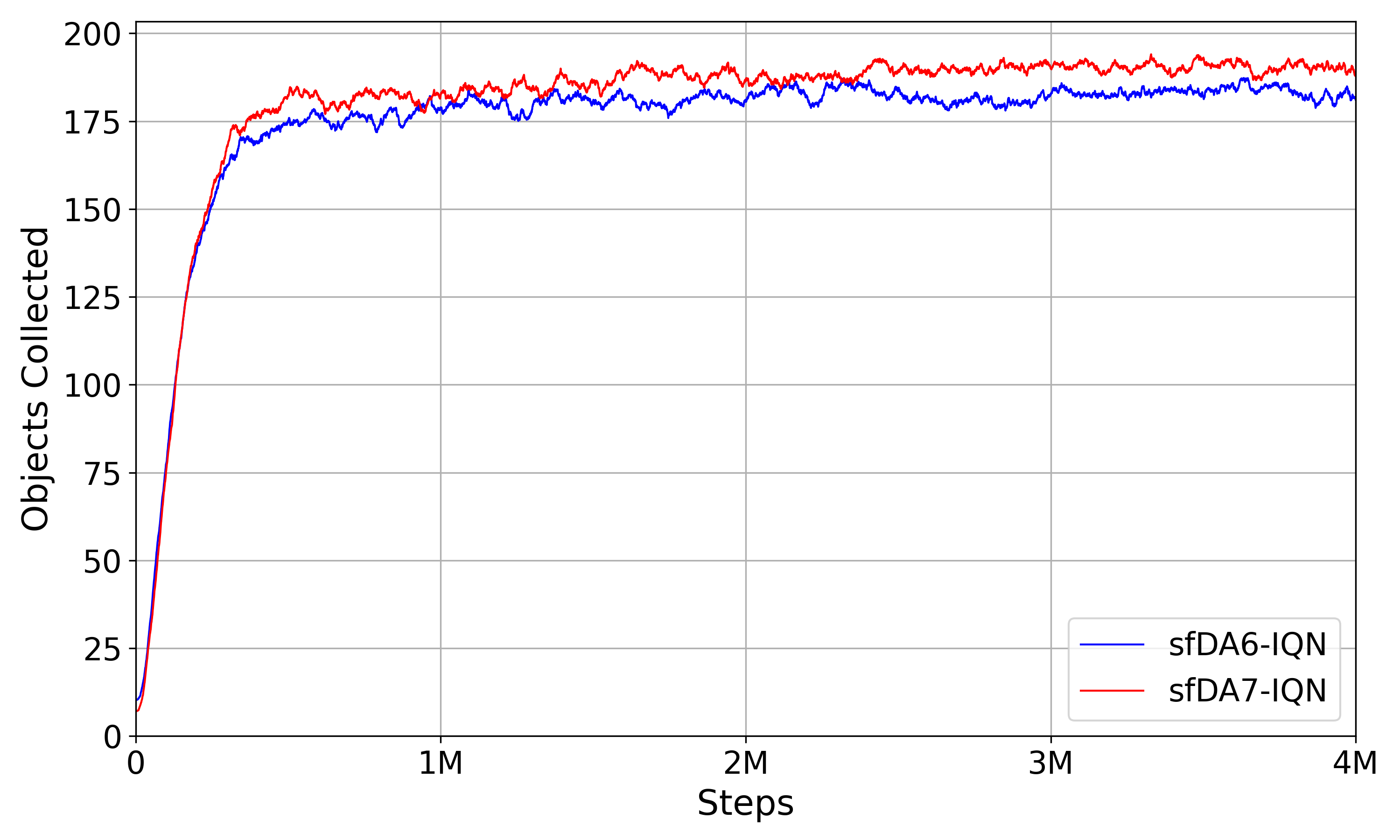}
  \caption{Learning curve of the average objects collected}
  \label{fig:curve_object}
\end{figure}

\begin{table}
    \centering
    \caption{Performance during training phase}
    \label{tab:training_reward_object}
    \begin{tabular}{lrr}
        \toprule
           Method & Reward & Objects collected \\
        \midrule
        sfDA6-IQN & 173.476 & 183.545 \\
        sfDA7-IQN & 179.124 & 190.513 \\
        Improvement ratio & 3.26\% & 3.80\% \\
        \bottomrule
    \end{tabular}
\end{table}

\subsection{Performance Comparison}
\subsubsection{Learning Speed and Performance}
First, we trained the baseline and proposed networks in agents and compared the learning speed and efficiency because additional matrices (ADCs) were fed to the networks in the proposed method, potentially reducing the learning efficiency. The resulting learning curves are presented in Fig.~\ref{fig:curve_object}, which plots the exponential moving average (EMA) of the objects collected in independent experimental runs with three random seeds, where the smoothing factor of the EMA was set to $0.99$. Table~\ref{tab:training_reward_object} lists the average rewards and numbers of collected objects per episode between $3$ M and $4$ M steps.
\par

\begin{figure}[t]
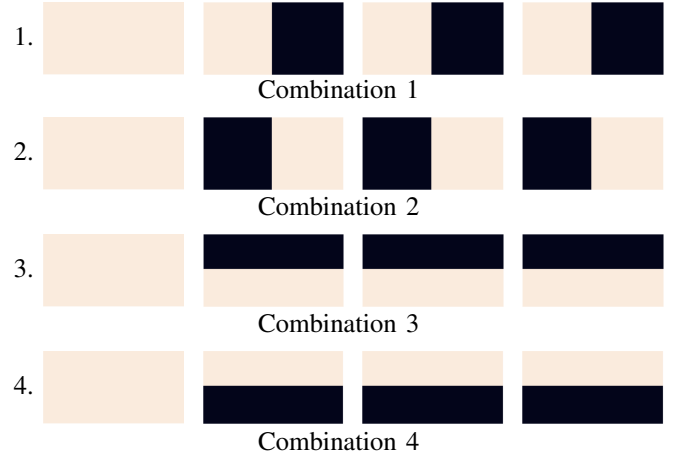

    \centering
    1. \begin{minipage}{0.21\linewidth}
        \includegraphics[width=\linewidth]{figures/entire.PNG}
    \end{minipage}
    \hfil
    \begin{minipage}{0.21\linewidth}
        \includegraphics[width=\linewidth]{figures/half0.PNG}
    \end{minipage}
    \hfil
    \begin{minipage}{0.21\linewidth}
        \includegraphics[width=\linewidth]{figures/half0.PNG}
    \end{minipage}
    \hfil
    \begin{minipage}{0.21\linewidth}
        \includegraphics[width=\linewidth]{figures/half0.PNG}
    \end{minipage}\\[2pt]
    Combination 1\\[7pt]
    2. \begin{minipage}{0.21\linewidth}
        \includegraphics[width=\linewidth]{figures/entire.PNG}
    \end{minipage}
    \hfil
    \begin{minipage}{0.21\linewidth}
        \includegraphics[width=\linewidth]{figures/half1.PNG}
    \end{minipage}
    \hfil
    \begin{minipage}{0.21\linewidth}
        \includegraphics[width=\linewidth]{figures/half1.PNG}
    \end{minipage}
    \hfil
    \begin{minipage}{0.21\linewidth}
        \includegraphics[width=\linewidth]{figures/half1.PNG}
    \end{minipage}\\[3pt]
    Combination 2\\[7pt]
    3. \begin{minipage}{0.21\linewidth}
        \includegraphics[width=\linewidth]{figures/entire.PNG}
    \end{minipage}
    \hfil
    \begin{minipage}{0.21\linewidth}
        \includegraphics[width=\linewidth]{figures/half2.PNG}
    \end{minipage}
    \hfil
    \begin{minipage}{0.21\linewidth}
        \includegraphics[width=\linewidth]{figures/half2.PNG}
    \end{minipage}
    \hfil
    \begin{minipage}{0.21\linewidth}
        \includegraphics[width=\linewidth]{figures/half2.PNG}
    \end{minipage}\\[3pt]
    Combination 3\\[7pt]
    4. \begin{minipage}{0.21\linewidth}
        \includegraphics[width=\linewidth]{figures/entire.PNG}
    \end{minipage}
    \hfil
    \begin{minipage}{0.21\linewidth}
        \includegraphics[width=\linewidth]{figures/half3.PNG}
    \end{minipage}
    \hfil
    \begin{minipage}{0.21\linewidth}
        \includegraphics[width=\linewidth]{figures/half3.PNG}
    \end{minipage}
    \hfil
    \begin{minipage}{0.21\linewidth}
        \includegraphics[width=\linewidth]{figures/half3.PNG}
    \end{minipage}\\[3pt]
    Combination 4
    \caption{DC combinations}
    \label{fig:exp1_DC}
\end{figure}

\begin{table}
    \centering
    \caption{Performance during execution phase}
    \label{tab:reward_object}
    \begin{tabular}{lrr}
        \toprule
           Method & Reward & Objects collected \\
        \midrule
        sfDA6-IQN & 133.895 & 135.357 \\
        sfDA7-IQN & 142.944 & 146.188 \\
        Improvement ratio & 6.76\% & 8.00\% \\
        \bottomrule
    \end{tabular}
\end{table}

\subsubsection{Performance with Biased DCs in Execution Phase}
We analyzed the performance and controllability of DCs and ADCs during execution by providing combinations of biased DCs. In this experiment, one of the four pattern combinations, as shown in Fig.~\ref{fig:exp1_DC}, was selected per experimental run, and each DC in the selected combination was sequentially provided to the agents (totaling 16 cases from combinations of four patterns and four agents). These four DC combinations indicate that one agent can collect objects anywhere, while the other three agents are directed to work in the same half of the area. This indicates a biased direction. Our interest lies in how the unrestricted agent in each episode moves to collect objects to compensate for the biased movements of others.
\par

\begin{figure}[b]
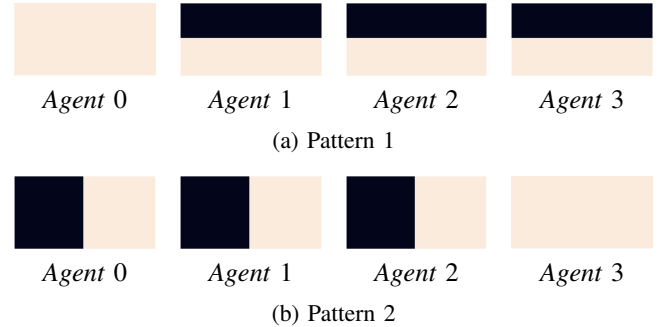

  \centering
  \begin{minipage}{\linewidth}
    \centering
  \begin{minipage}{0.21\linewidth}
    \centering
    \includegraphics[width=\linewidth]{figures/entire.PNG} \\
    \Agent~0
  \end{minipage}
    \hfil
  \begin{minipage}{0.21\linewidth}
    \centering
    \includegraphics[width=\linewidth]{figures/half2.PNG} \\
    \Agent~1
  \end{minipage}
    \hfil
  \begin{minipage}{0.21\linewidth}
    \centering
    \includegraphics[width=\linewidth]{figures/half2.PNG} \\
    \Agent~2
  \end{minipage}
    \hfil
  \begin{minipage}{0.21\linewidth}
    \centering
    \includegraphics[width=\linewidth]{figures/half2.PNG} \\
    \Agent~3
  \end{minipage}
  \subcaption{Pattern~1}\label{fig:exp1_p1}
  \end{minipage}
  \\[8pt]
  \begin{minipage}{\linewidth}
    \centering
    \begin{minipage}{0.21\linewidth}
        \centering
        \includegraphics[width=\linewidth]{figures/half1.PNG} \\
        \Agent~0
    \end{minipage}
    \hfil
    \begin{minipage}{0.21\linewidth}
        \centering
        \includegraphics[width=\linewidth]{figures/half1.PNG} \\
        \Agent~1
    \end{minipage}
    \hfil
    \begin{minipage}{0.21\linewidth}
        \centering
        \includegraphics[width=\linewidth]{figures/half1.PNG} \\
        \Agent~2
    \end{minipage}
    \hfil
    \begin{minipage}{0.21\linewidth}
        \centering
        \includegraphics[width=\linewidth]{figures/entire.PNG} \\
        \Agent~3
    \end{minipage}
    \subcaption{Pattern~2}\label{fig:exp1_p2}
  \end{minipage}
  \caption{Combinations of DCs (biased)}\label{fig:exp1_patterns}
\end{figure}

\begin{figure*}
  \centering
\begin{minipage}{0.45\linewidth}
  \centering
  \includegraphics[width=0.8\linewidth]{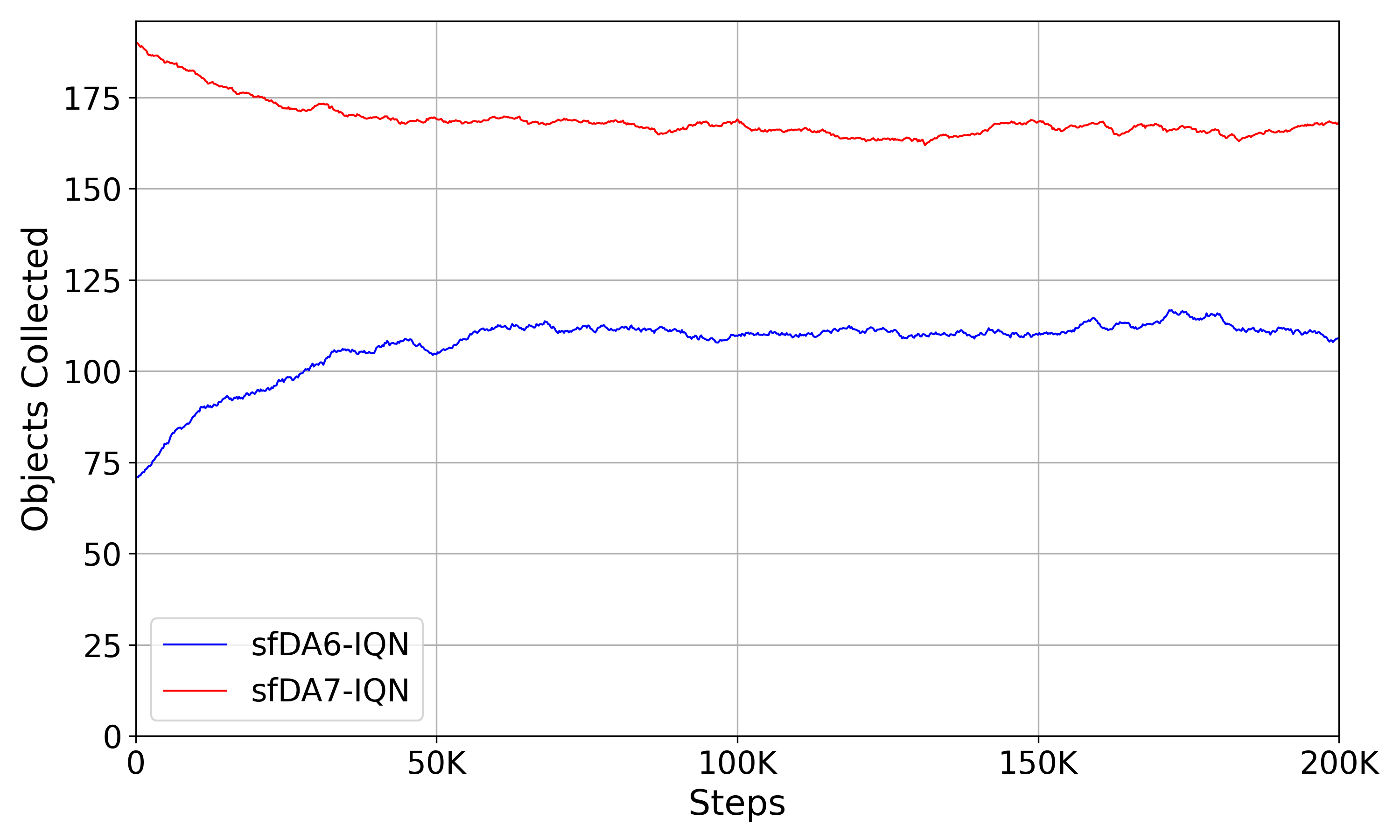}
  \subcaption{Pattern~1}
  \label{fig:exp1_p1_object}
\end{minipage}
\hfil
\begin{minipage}{0.45\linewidth}
  \centering
  \includegraphics[width=0.8\linewidth]{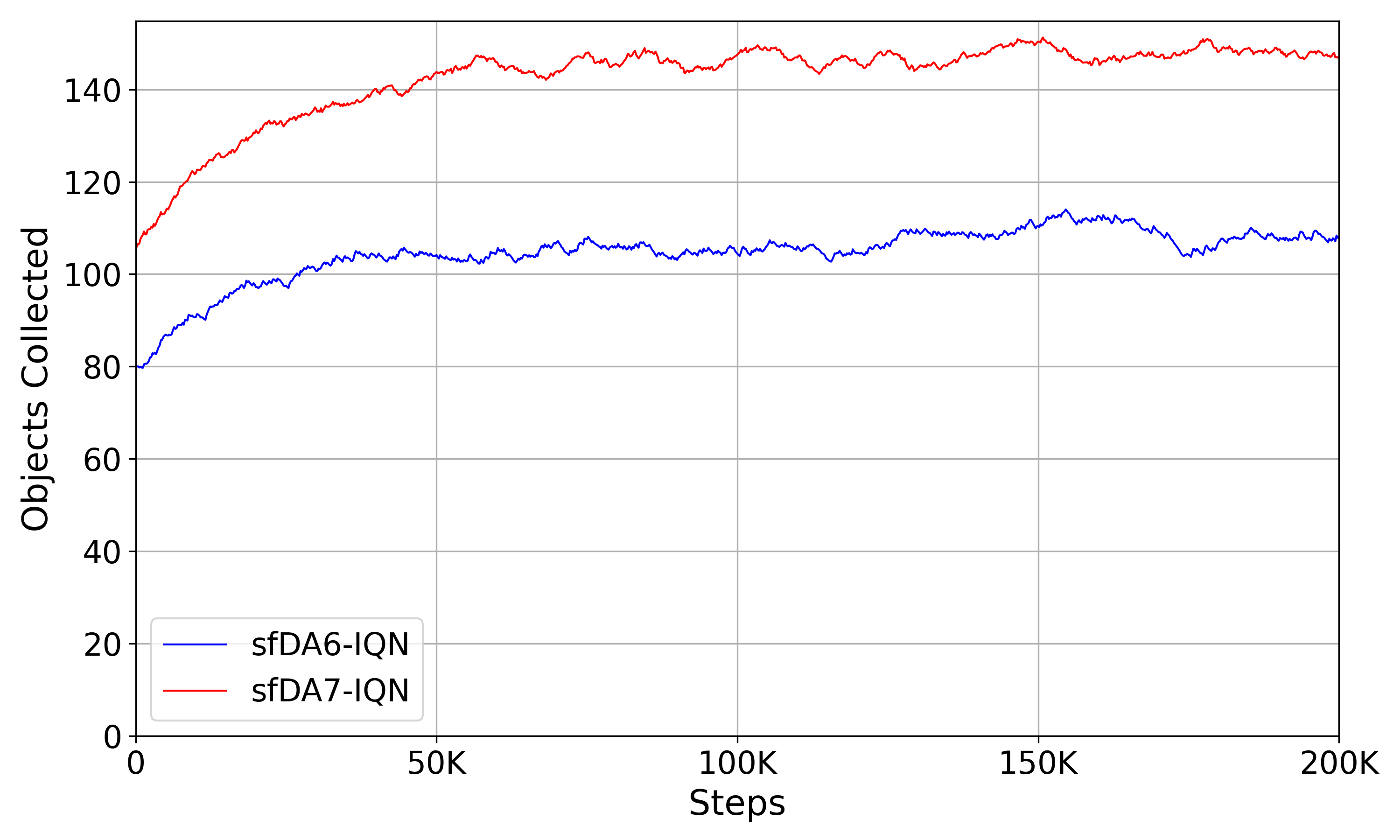}
  \subcaption{Pattern~2}
  \label{fig:exp1_p2_object}
\end{minipage}
  \caption{Numbers of collected objects}
\end{figure*}

The results obtained using the three random seeds are listed in Table~\ref{tab:reward_object}. The proposed method achieved an $8.00\%$ improvement in the number of collected objects and a $6.76\%$ improvement in the rewards compared with the baseline method. This suggests that the agent without a specific direction attempted to collect more objects, considering the movement areas of the other agents based on the DCs provided to them.
\par

\begin{figure*}
  \begin{minipage}{\linewidth}
    \begin{minipage}{0.19\linewidth}
        \centering
        \includegraphics[width=\linewidth]{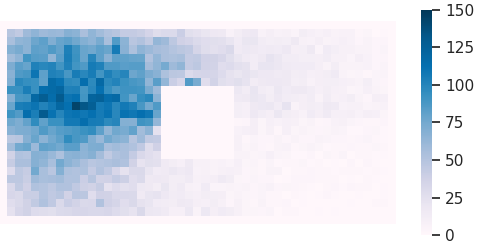}
        \footnotesize \Agent~0
    \end{minipage}
    \hfill
    \begin{minipage}{0.19\linewidth}
        \centering
        \includegraphics[width=\linewidth]{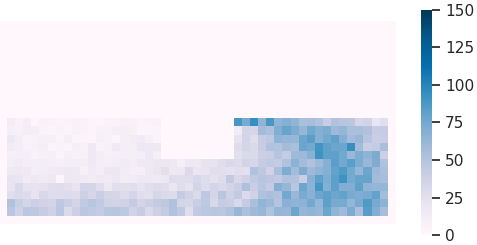}
        \footnotesize \Agent~1
    \end{minipage}
    \hfill
    \begin{minipage}{0.19\linewidth}
        \centering
        \includegraphics[width=\linewidth]{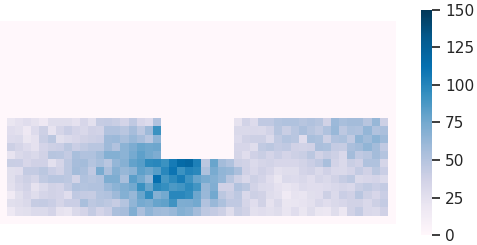}
        \footnotesize \Agent~2
    \end{minipage}
    \hfill
    \begin{minipage}{0.19\linewidth}
        \centering
        \includegraphics[width=\linewidth]{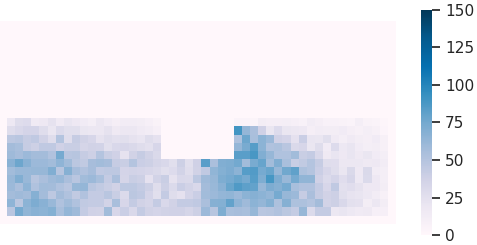}
        \footnotesize \Agent~3
    \end{minipage}
    \subcaption{sfDA6-IQN}
  \end{minipage}
  \\[5pt]
  \begin{minipage}{\linewidth}
    \begin{minipage}{0.19\linewidth}
        \centering
        \includegraphics[width=\linewidth]{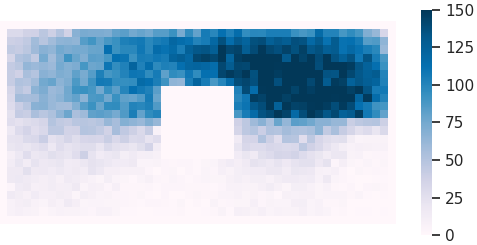}
        \footnotesize \Agent~0
    \end{minipage}
    \hfill
    \begin{minipage}{0.19\linewidth}
        \centering
        \includegraphics[width=\linewidth]{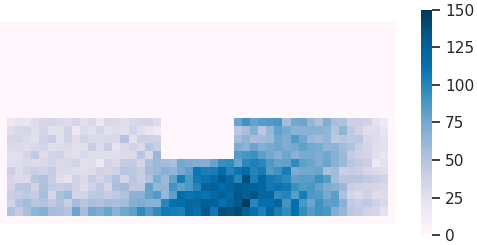}

        \footnotesize \Agent~1
    \end{minipage}
    \hfill
    \begin{minipage}{0.19\linewidth}
        \centering
        \includegraphics[width=\linewidth]{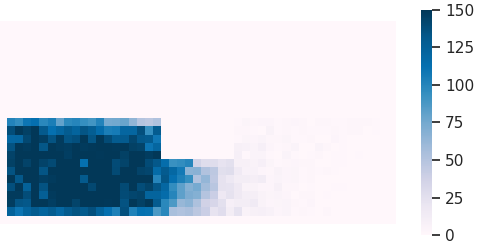}
        \footnotesize \Agent~2
    \end{minipage}
    \hfill
    \begin{minipage}{0.19\linewidth}
        \centering
        \includegraphics[width=\linewidth]{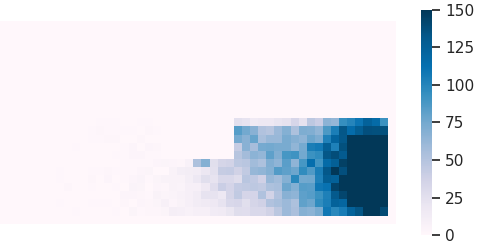}
        \footnotesize \Agent~3
    \end{minipage}
    \subcaption{sfDA7-IQN}
  \end{minipage}
  \caption{Object collection locations and counts (Pattern~1)}
  \label{fig:exp1_p1_heatmap}
\end{figure*}

\begin{figure*}
  \centering
  \begin{minipage}{\linewidth}
    \begin{minipage}{0.19\linewidth}
        \centering
        \includegraphics[width=\linewidth]{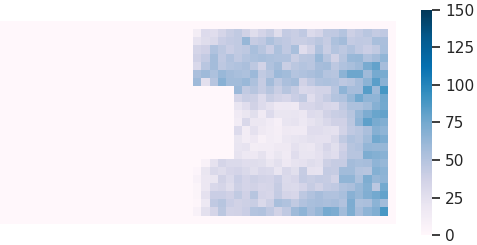}
        \footnotesize \Agent~0
    \end{minipage}
    \hfill
    \begin{minipage}{0.19\linewidth}
        \centering
        \includegraphics[width=\linewidth]{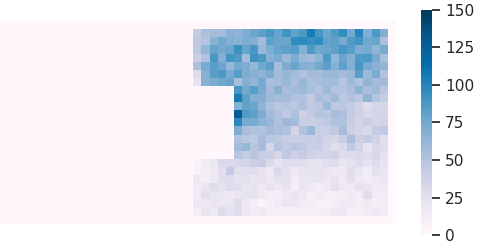}
        \footnotesize \Agent~1
    \end{minipage}
    \hfill
    \begin{minipage}{0.19\linewidth}
        \centering
        \includegraphics[width=\linewidth]{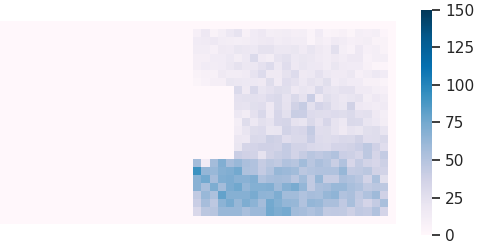}
        \footnotesize \Agent~2
    \end{minipage}
    \hfill
    \begin{minipage}{0.19\linewidth}
        \centering
        \includegraphics[width=\linewidth]{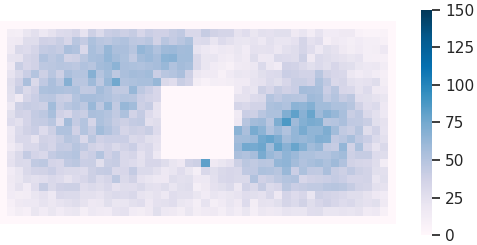}
        \footnotesize \Agent~3
    \end{minipage}
    \subcaption{sfDA6-IQN}
    \end{minipage}
  \\[5pt]
  \begin{minipage}{\linewidth}
    \begin{minipage}{0.19\linewidth}
        \centering
        \includegraphics[width=\linewidth]{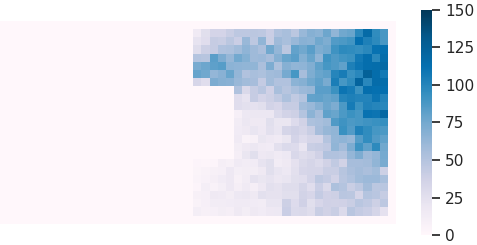}
        \footnotesize \Agent~0
    \end{minipage}
    \hfill
    \begin{minipage}{0.19\linewidth}
        \centering
        \includegraphics[width=\linewidth]{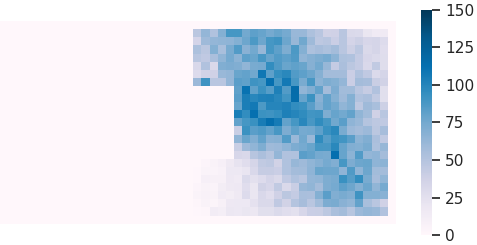}
        \footnotesize \Agent~1
    \end{minipage}
    \hfill
    \begin{minipage}{0.19\linewidth}
        \centering
        \includegraphics[width=\linewidth]{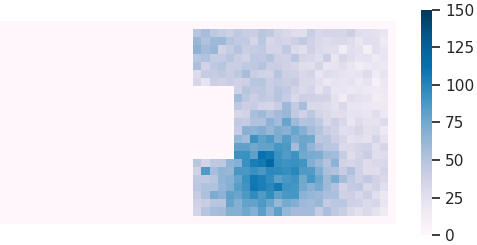}
        \footnotesize \Agent~2
    \end{minipage}
    \hfill
    \begin{minipage}{0.19\linewidth}
        \centering
        \includegraphics[width=\linewidth]{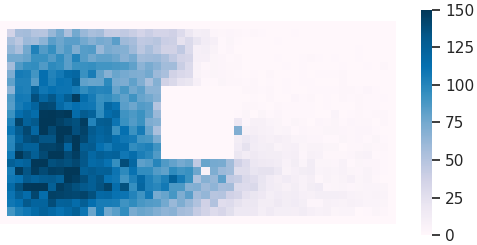}
        \footnotesize \Agent~3
    \end{minipage}
  \subcaption{sfDA7-IQN}
  \end{minipage}
  \caption{Object collection locations and counts (Pattern~2)}
  \label{fig:exp1_p2_heatmap}
\end{figure*}

\subsection{Detailed Analysis of Activity Locations and Counts}
\subsubsection{Biased Directions}\label{sec:biasd-DCs}
To validate our proposed method, we analyzed where agents collected objects and the number of objects using both methods. Two experimental runs were selected, and heatmaps were generated to illustrate the results. In the first instance, four DCs called {\em Pattern~1} were provided to the agents, as shown in Fig.~\ref{fig:exp1_p1}, where \Agents~1, 2, and 3 were directed to the lower half of the environment, whereas Agent~0 was not directed explicitly. In the second instance, DCs called {\em Pattern~2} were provided (Fig.~\ref{fig:exp1_p2}), where \Agents~0, 1, and 2 were directed to work in the right half, whereas \Agent~3 was not.
\par

First, we confirmed the performances of both methods. The number of objects collected by agents with sfDA6-IQN and sfDA7-IQN over time is plotted in Figs.~\ref{fig:exp1_p1_object} (Pattern~1) and \ref{fig:exp1_p2_object} (Pattern~2). These figures indicate that the agents with sfDA7-IQN outperformed those with sfDA6-IQN. Note that although the curves increased or decreased before 50 K steps, these fluctuations were caused by the initial data due to the exponential moving average calculation; therefore, we focused on the values after 100 K steps.
\par

\begin{figure}[t]
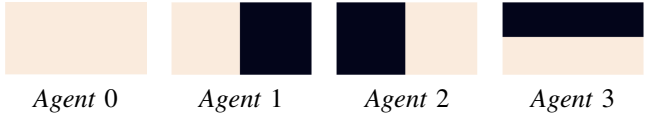

    \centering
    \begin{minipage}{0.21\linewidth}
        \centering
        \includegraphics[width=\linewidth]{figures/entire.PNG} \\
        \Agent~0
    \end{minipage}
    \hfil
    \begin{minipage}{0.21\linewidth}
        \centering
        \includegraphics[width=\linewidth]{figures/half0.PNG} \\
        \Agent~1
    \end{minipage}
    \hfil
    \begin{minipage}{0.21\linewidth}
        \centering
        \includegraphics[width=\linewidth]{figures/half1.PNG} \\
        \Agent~2
    \end{minipage}
    \hfil
    \begin{minipage}{0.21\linewidth}
        \centering
        \includegraphics[width=\linewidth]{figures/half2.PNG} \\
        \Agent~3
    \end{minipage}
    \caption{Combinations of DCs (Exp 2)}
    \label{fig:exp2}
\end{figure}

\begin{figure}[t]
  \centering
  \includegraphics[width=0.75\linewidth]{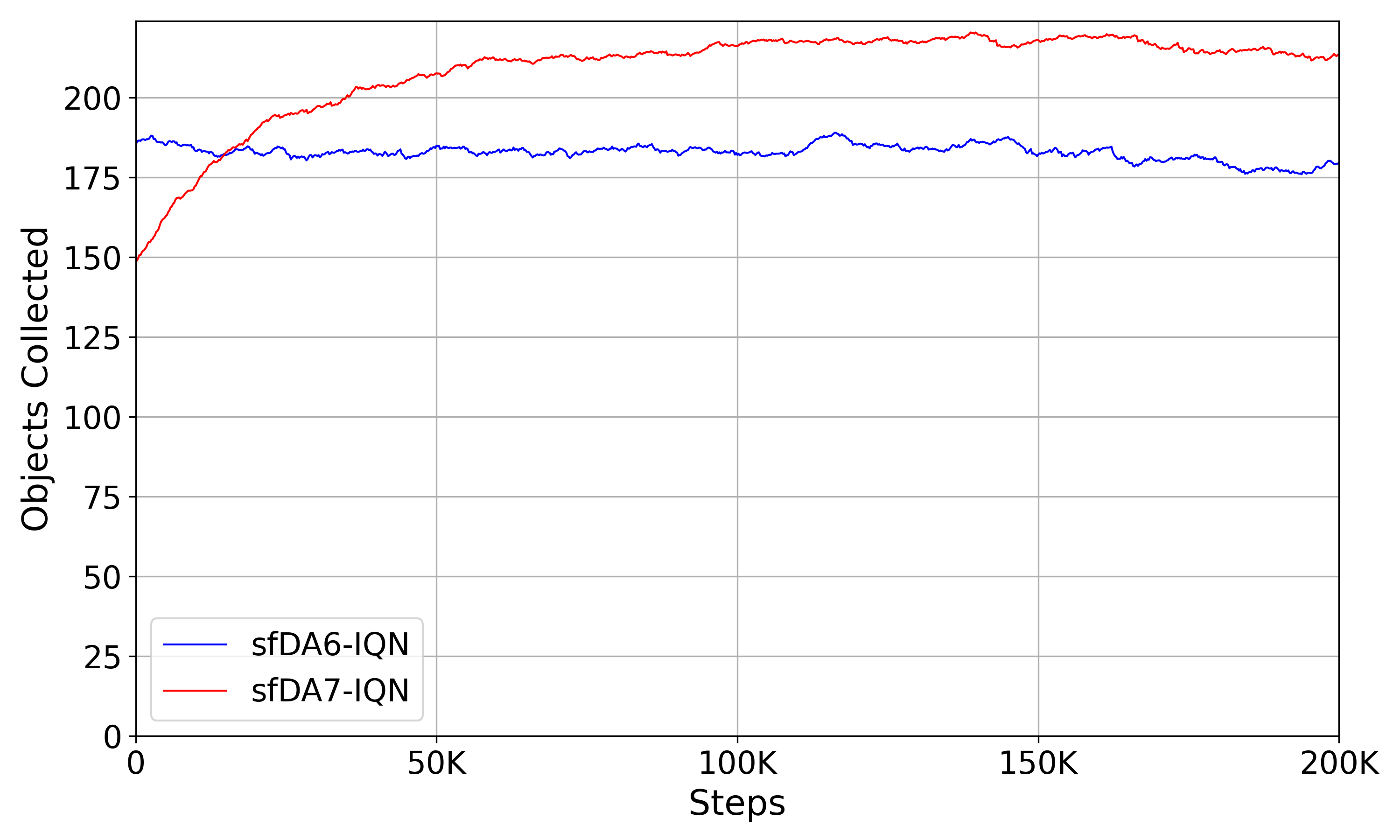}
  \caption{Objects collected (Exp 2)}
  \label{fig:exp2_object}
\end{figure}

\begin{figure*}[!b]
  \centering
  \begin{minipage}{\linewidth}
    \begin{minipage}{0.19\linewidth}
        \centering
        \includegraphics[width=\linewidth]{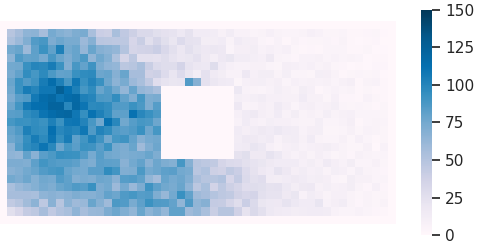}
        \footnotesize \Agent~0
    \end{minipage}
    \hfill
    \begin{minipage}{0.19\linewidth}
        \centering
        \includegraphics[width=\linewidth]{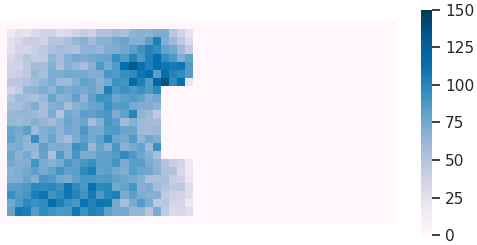}
        \footnotesize \Agent~1
    \end{minipage}
    \hfill
    \begin{minipage}{0.19\linewidth}
        \centering
        \includegraphics[width=\linewidth]{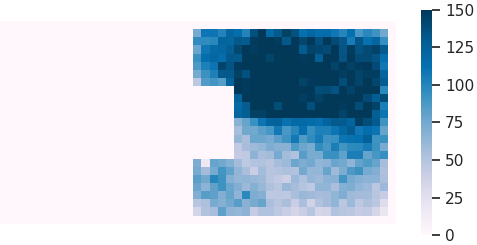}
        \footnotesize \Agent~2
    \end{minipage}
    \hfill
    \begin{minipage}{0.19\linewidth}
        \centering
        \includegraphics[width=\linewidth]{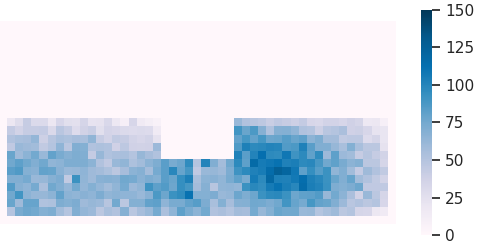}
        \footnotesize \Agent~3
    \end{minipage}
    \subcaption{sfDA6-IQN}
  \end{minipage}
  \\[8pt]
  \begin{minipage}{\linewidth}
    \begin{minipage}{0.19\linewidth}
        \centering
        \includegraphics[width=\linewidth]{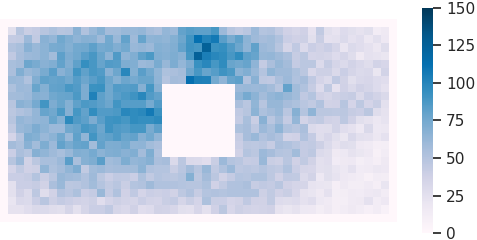}
        \footnotesize \Agent~0
    \end{minipage}
    \hfill
    \begin{minipage}{0.19\linewidth}
        \centering
        \includegraphics[width=\linewidth]{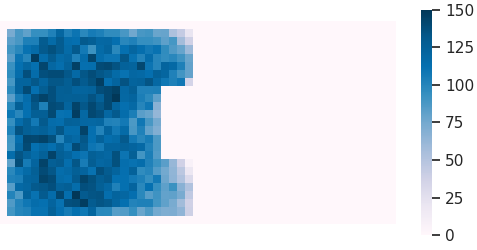}
        \footnotesize \Agent~1
    \end{minipage}
    \hfill
    \begin{minipage}{0.19\linewidth}
        \centering
        \includegraphics[width=\linewidth]{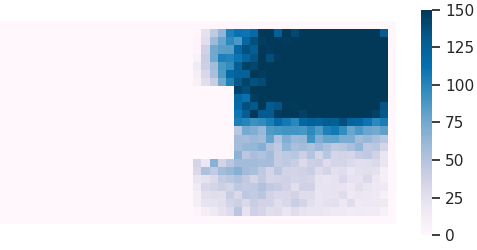}
        \footnotesize \Agent~2
    \end{minipage}
    \hfill
    \begin{minipage}{0.19\linewidth}
        \centering
        \includegraphics[width=\linewidth]{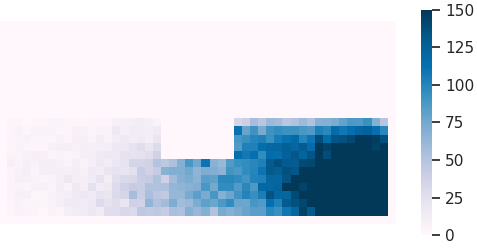}
        \footnotesize \Agent~3
    \end{minipage}
  \subcaption{sfDA7-IQN}
  \end{minipage}
  \caption{Object collection locations (Exp 2)}
  \label{fig:exp2_heatmap}
\end{figure*}

Figures~\ref{fig:exp1_p1_heatmap} and \ref{fig:exp1_p2_heatmap} show heatmaps indicating where agents collected objects when Patterns~1 and 2 were provided. In the first case (Fig.~\ref{fig:exp1_p1_heatmap}), {\em Agents}~1, 2, and 3 collected objects in the lower-half area with both the sfDA6-IQN and sfDA7-IQN methods by following the DCs. Agent~0 with sfDA6-IQN collected objects without considering the DCs of other agents, redundantly collecting in the lower-left area while ignoring the upper-right area. In contrast, Agent~0 with sfDA7-IQN collected objects mainly in the upper-half area to compensate for the biased DCs provided to the other agents. Looking at the heatmaps for \Agents~1, 2, and 3, agents with sfDA7-IQN collected more objects (darker working areas), likely because they considered other agents' DCs and learned the optimal movement during training.
\par

In the second experiment, the agents with sfDA7-IQN collected more objects, as shown in Fig.~\ref{fig:exp1_p2_heatmap}. \Agents~0, 1, and 2 with sfDA6-IQN moved around in the right-half area. Similarly, \Agents~0, 1, and 2 with sfDA7-IQN moved to the right-half area, showing a slightly darker color in Fig.~\ref{fig:exp1_p2_heatmap}. \Agent~3 with sfDA6-IQN redundantly collected objects throughout the environment. \Agent~3 with sfDA7-IQN collected objects mainly in the left-half area to compensate for other agents, thereby increasing the overall efficiency through complementary coordination.
\par

\subsubsection{Unbiased Directions}
Another experiment was conducted using the DC combination pattern shown in Fig.~\ref{fig:exp2}. These DCs cover the environment; however, the agents collected objects more frequently in the lower half. The number of objects collected over time is presented in Fig.~\ref{fig:exp2_object}. After $100K$ steps, agents with sfDA7-IQN collected approximately $18.1\%$ more objects than those with sfDA6-IQN.
\par

Heatmaps indicating where agents collected the objects are shown in Fig.~\ref{fig:exp2_heatmap}. Similar to Figs.~\ref{fig:exp1_p1_heatmap} and \ref{fig:exp1_p2_heatmap}, \Agents~1, 2, and 3 collected objects only in areas directed by DCs, and agents with sfDA7-IQN collected more objects because their heatmaps were darker than those with sfDA6-IQN. Because \Agent~0 with sfDA6-IQN could not consider DCs provided to other agents, its collected area was inconsistent with the other agents' heatmaps. However, \Agent~0 with sfDA7-IQN collected objects throughout the environment, with more objects in the upper half. The difference between the upper and lower halves was not apparent compared to the results with biased DCs. The values in the ADC matrix generated from the DCs of \Agents~1, 2, and 3 were $1$ or $2$. This shows that the entire area was covered, and the difference was $1$ ($=2-1$), which is smaller than that in Section~\ref{sec:biasd-DCs}, which was $3$ ($=3-0$). Therefore, \Agent~0 collected objects throughout the entire area, not just in the upper half.
\par

\subsubsection{Remark: Other Usages of DCs and ADCs}
Human managers may intend that an agent provided with an entire DC acts solely on that DC without considering the DCs of other agents. This can be achieved by setting all ADC matrix elements to values such as $N-1$ and $0$. DCs can also reinforce behavior in specific areas. In the learning experiment described herein, random DCs provided to the agents caused them to move evenly around all areas. Alternatively, agents can be directed to frequently move around specific areas, such as the lower half, without increasing the number of objects or rewards using DC combinations to direct such activities.
\par

\section{Conclusion}
We proposed sfDA7-X to establish agent controllability after learning using MADRL. This method extends sfDA6-X~\cite{motokawa2023strategy} by introducing ADCs to express DCs provided to other agents in an aggregated form and feeding them into agents' networks. ADCs enable agents to consider the activities of others to complement areas in which other agents collect no objects or visit infrequently. Experiments showed that agents with sfDA7-X outperformed those with sfDA6-X and achieved more reasonable activities without specific directions. The learning efficiency remained comparable to that of sfDA6-X despite additional network inputs. Future work will include extending the proposed method to improve learning in more complex environments using curriculum learning based on DCs and ADCs.
\par

\bibliographystyle{IEEEtranS}
\bibliography{IEEEabrv,references}

\end{document}